\documentclass[aps,prl,twocolumn,superscriptaddress,groupedaddress,nofootinbib]{revtex4}
\usepackage{amsmath,amsfonts,amssymb,array}
\usepackage{bm,bbm,bbm,bbold}
\usepackage{enumerate}
\usepackage{dsfont}
\usepackage{float}
\usepackage{graphicx}
\usepackage{longtable}
\usepackage{multirow}
\usepackage{slashed,subfigure}
\usepackage{xcolor}
\usepackage{physics}
\usepackage[colorlinks = true,linkcolor = blue]{hyperref} 
\definecolor{darkgreen}{rgb}{0.0, 0.5, 0.0}
\hypersetup{unicode=true,pdftoolbar=true,pdfmenubar=true,
  pdffitwindow=false,pdfstartview={FitH},
  pdftitle={NNforDEsPT},pdfauthor={Piscopo,Spannowsky,Waite},
  pdfkeywords={Neural Networks}{Phase Transitions}{Differential Equations},
  pdfnewwindow=true,colorlinks=true,
  linkcolor=blue,citecolor=purple,filecolor=magenta,urlcolor=cyan}

\begin{document}
\leftline{}
\rightline{IPPP/19/11}

\title{Solving differential equations with neural networks: \\ Applications to the calculation of cosmological phase transitions}

\author{Maria Laura Piscopo}  
\email{maria.l.piscopo@durham.ac.uk}
\affiliation{Institute for Particle Physics Phenomenology, Department of Physics, Durham University, Durham DH1 3LE, U.K.}

\author{Michael Spannowsky}  
\email{michael.spannowsky@durham.ac.uk}
\affiliation{Institute for Particle Physics Phenomenology, Department of Physics, Durham University, Durham DH1 3LE, U.K.}

\author{Philip Waite}  
\email{p.a.waite@durham.ac.uk}
\affiliation{Institute for Particle Physics Phenomenology, Department of Physics, Durham University, Durham DH1 3LE, U.K.}

\date{\today}

\begin{abstract}
Starting from the observation that artificial neural networks are uniquely suited to solving optimisation problems, and most physics problems can be cast as an optimisation task, we introduce a novel way of finding a numerical solution to wide classes of differential equations. We find our approach to be very flexible and stable without relying on trial solutions, and applicable to ordinary, partial and coupled differential equations. 
We apply our method to the calculation of tunnelling profiles for cosmological phase transitions, which is a problem of relevance for baryogenesis and stochastic gravitational wave spectra. Comparing our solutions with publicly available codes which use numerical methods optimised for the calculation of tunnelling profiles, we find our approach to provide at least as accurate results as these dedicated differential equation solvers, and for some parameter choices even more accurate and reliable solutions. In particular, we compare the neural network approach with two publicly available profile solvers, \texttt{CosmoTransitions} and \texttt{BubbleProfiler}, and give explicit examples where the neural network approach finds the correct solution while dedicated solvers do not. We point out that this approach of using artificial neural networks to solve equations is viable for any problem that can be cast into the form $\mathcal{F}(\vec{x})=0$, and is thus applicable to various other problems in perturbative and non-perturbative quantum field theory.

\end{abstract}

\maketitle

\section{Introduction}\label{sec:Intro}

A neural network is an algorithm designed to perform an optimisation procedure, where the loss function provides a measure of the performance of the optimisation. Thus, if a physics problem can be cast into the form $\mathcal{F}(\vec{x})=0$, then its solution can be calculated by minimising the loss function of a neural network. While this approach is applicable to any function $\mathcal{F}$, we attempt to apply this observation to the solution of differential equations and to the non-perturbative calculation of tunnelling rates of electroweak phase transitions.

Solving differential equations is a profound problem, relevant for all areas of theoretical physics. For large classes of differential equations, analytic solutions cannot be found. Thus, numerical or approximative methods are needed to solve them. Standard methods to solve differential equations numerically include the Runge-Kutta method, linear multistep methods, finite-element or finite-volume methods, and spectral methods \cite{EncyMath}. We instead propose a novel approach to solving differential equations using artificial neural networks.

In recent years, machine-learning algorithms have become increasingly popular in extracting correlations in high-dimensional parameter spaces. Even for a small number of dimensions, e.g. $n_\mathrm{dim} \geq 3$, it becomes very difficult to visualise data such that a human can extract correlations to a high degree of accuracy. Machine-learning algorithms, and in particular neural networks, prove to be faster and more precise and allow a parametric improvement of the precision in how well the region of interest is interpolated. As a result, various neural network architectures have been designed, e.g. convolutional neural networks, recurrent neural networks, deep neural networks, etc.,  to perform various increasingly complex tasks.

In particle physics such tasks include the classification of signal-to-background events based on event selection cuts \cite{McGlone:1356221,Chatrchyan:2013kha,Aaboud:2017rss,Sirunyan:2018hbu}, or the classification of complex objects such as jets, according to the image their radiation imprints in the detector \cite{Almeida:2015jua, deOliveira:2015xxd, Barnard:2016qma,Komiske:2016rsd, Lenz:2017lqo,Metodiev:2017vrx,Kasieczka:2017nvn,Butter:2017cot, Chang:2017kvc}. In other applications neural networks are used to regress between data points \cite{Ball:2008by,Louppe:2016ylz, DAgnolo:2018cun,Englert:2018cfo,Brehmer:2018kdj} or are trained on a well-known sample to identify outliers or anomalies \cite{Farina:2018fyg,Hajer:2018kqm,Collins:2018epr}. However, in all aforementioned applications that can be characterised as classification and regression, the neural network is applied to an output sample, trying to extract information on the parameters that determine the input. In particle physics that would mean to analyse the radiation profile as recorded by a particle detector to learn the parameters of the underlying model, e.g. the Standard Model. Input and output are connected through quantum field theory, i.e. a non-trivial set of differential and integral equations.

We propose to use these powerful artificial neural network algorithms in a different way, namely to directly find solutions to differential equations. We then apply these methods to calculate the solution of the non-perturbative quantum-field-theoretical description of tunnelling processes for electroweak phase transitions. The fast and reliable calculation of tunnelling rates of the electroweak phase transitions within and in extensions of the Standard Model is of importance to assessing if the model allows for a strong first-order phase transition during the evolution of the early Universe. This could explain baryogenesis \cite{Sakharov:1967dj, Kuzmin:1985mm} in such a model as the source of matter-antimatter asymmetry in the Universe, and further lead to a stochastic gravitational wave signal which could potentially be measured at future gravitational wave experiments \cite{Kosowsky:1992rz, Grojean:2006bp}, e.g. eLISA \cite{Caprini:2015zlo}.

The universal approximation theorem \cite{Hornik:1989, Hornik:1991} allows us to expect a neural network to perform well in solving complicated mathematical expressions. It states that an artificial neural network containing a single hidden layer can approximate any arbitrarily complex function with enough neurons. We make use of this property by proposing a neural network model where the output of the network alone solves the differential equation, subject to its boundary conditions. In contrast to previous approaches \cite{Lee:1990,Maede:1994,Maede:19942,Lagaris:1997at,Mall:2013,iftikhar:2014} where the neural network is part of a full trial solution which is fixed to satisfy the boundary conditions, our approach includes the boundary conditions as additional terms in the loss function. The derivatives of the network output with respect to its inputs are calculated and passed to the loss function, and the network is optimised via backpropagation to regress to the solution of the differential equation. The network then gives a fully differentiable function which can be evaluated at any point within the training domain, and in some cases, extrapolated to further points (although we do not explore the extrapolation performance here).

We will begin by describing the method in detail and showcasing how it can be used to solve differential equations of varying complexity, before applying it to the calculation of cosmological phase transitions.

\section{The Method}\label{sec:method}

\subsection{Design of the network and optimisation}

We consider an artificial feedforward neural network (NN) with $n$ inputs, $m$ outputs and a single hidden layer with $k$ units. The outputs of the network, $N_m(\vec{x},\{w,\vec{b}\})$, can be written as,
\begin{equation}
\label{eq:network_equation}
N_m(\vec{x},\{w,\vec{b}\}) = \sum_{k,n} w^f_{mk} g(w^h_{kn}x_n+b^h_k) + b^f_m~,
\end{equation}
where the activation function $g: \mathbf{R}^k \mapsto \mathbf{R}^k$ is applied element-wise to each unit, and $h$ and $f$ denote the hidden and final layers, respectively. We use a single neural network with $m$ outputs to predict the solutions to $m$ coupled differential equations, and for the case of one differential equation, we use $m=1$.

A set of $m$ coupled $j$th order differential equations can be expressed in the general form,
\begin{equation}
\label{eq:diff_eq_general_form}
\mathcal{F}_m(\vec{x},\phi_m(\vec{x}),\nabla \phi_m(\vec{x}),\cdots, \nabla^{j} \phi_m(\vec{x})) = 0~,
\end{equation}
with boundary or initial conditions imposed on the solutions $\phi_m(\vec{x})$.
Writing the differential equations in such a way allows us to easily convert the problem of finding a solution into an optimisation one. An approximate solution $\hat{\phi}_m(\vec{x})$ is one which approximately minimises the square of the left-hand side of Eq.~\eqref{eq:diff_eq_general_form}, and thus the analogy can be made to the loss function of a neural network. In previous approaches \cite{Lee:1990,Maede:1994,Maede:19942,Lagaris:1997at}, $\hat{\phi}_m(\vec{x})$ is a trial solution composed of two parts: one which satisfies the boundary conditions, and one which is a function of the output of a neural network and vanishes at the boundaries. However, this requires one to choose a special form of the trial solution which is dependent on the boundary conditions. Furthermore, for some configurations of boundary conditions, finding such a trial solution is a very complex task, e.g. in the case of phase transitions. Instead, we identify the trial solution with the network output, $\hat{\phi}_m(\vec{x}) \equiv N_m(\vec{x},\{w,\vec{b}\})$, and include the boundary conditions as extra terms in the loss function.
If the domain is discretised into a finite number of training points $\vec{x}^i$, then approximations to the solutions, $\hat{\phi}_m(\vec{x})$, can be obtained by finding the set of weights and biases, $\{w,\vec{b}\}$, such that the neural network loss function is minimised on the training points. For $i_{\mathrm{max}}$ training examples, the full loss function that we use is,
\begin{align}
\label{eq:general_loss_function}
\mathcal{L}(\{w,\vec{b}\}) & = \frac{1}{i_{\mathrm{max}}}\sum_{i,m} \hat{\mathcal{F}}_m(\vec{x}^i,\hat{\phi}_m(\vec{x}^i),\cdots, \nabla^{j} \hat{\phi}_m(\vec{x}^i))^2 \nonumber \\ 
&+ \sum_{\mathrm{B.C.}}(\nabla^{p}\hat{\phi}_m(\vec{x}_b)-K(\vec{x}_b))^2~,
\end{align}
where the second term represents the sum of the squares of the boundary conditions, defined at the boundaries $\vec{x}_b$.\footnote{Here, $p$ represents the order of derivative for which the boundary condition is defined, and $K$ is a function on the boundary. For example, for the condition $\frac{d}{dx}\phi(0)=1$ the second term would be $\left(\frac{d}{dx}\hat{\phi}(0)-1\right)^2$.} These can be Dirichlet or Neumann, or they can be initial conditions if defined at the initial part of the domain.

The problem is then to minimise $\mathcal{L}(\{w,\vec{b}\})$ by optimising the weights and biases in the network, for a given choice of network setup. To calculate the loss, it is necessary to compute the derivatives of the network output with respect to its input. Since each part of the network, including the activation functions, are differentiable, then the derivatives can be obtained analytically. Ref.~\cite{Lagaris:1997at} outlines how to calculate these derivatives. The optimisation can then proceed via backpropagation by further calculating the derivatives of the loss itself with respect to the network parameters. We use the \texttt{Keras} framework \cite{chollet2015keras} with a \texttt{TensorFlow} \cite{Abadi:2016kic} backend to implement the network and perform the optimisation of the loss function.

As with any neural network, the choice of hyperparameters will have an effect on the performance. For our setup, the important parameters are the number of hidden layers, the number of units in each hidden layer, the number of training points $\vec{x}^{(i)}$ (corresponding to the number of anchors in the discretisation of the domain of the differential equation), the activation function in each hidden layer, the optimisation algorithm, the learning rate, and the number of epochs the network is trained for. Furthermore, a choice must be made for the size of the domain that contains the points that the network is trained on, but this will usually be determined by the problem being solved.

In all the examples, we use the Adam optimiser \cite{Kingma:2014} with learning rate reduction on plateau---i.e. when the loss plateaus, the learning rate is reduced---and an initial learning rate of $0.01$. We find that the network is not susceptible to overfitting---the training points are chosen exactly from the domain that one is trying to find the solution to, and are not subject to statistical fluctuations; thus, finding a solution for which the loss at every training point is zero would not limit the generalisation of the solution to other points within the domain. Therefore, we use a large number of epochs such that the training loss becomes very small. For all examples we use a conservative number of $5\times10^4$ epochs. Furthermore, we use the entire set of training points in each batch so that the boundary conditions in the loss are included for each update of the network parameters. We also find that, in general, a single hidden layer with a small number of units [$\mathcal{O}(10)$] is sufficient to obtain very accurate solutions.

In order to assess and improve the stability and performance in certain cases, there are some additional technical methods which we employ beyond the basic setup. Firstly, the differentiability of the network solution allows us to calculate the differential contribution, $\hat{\mathcal{F}}$, to the loss across the entire training domain. This shows the degree of accuracy to which each part of the network solution satisfies the differential equation, and can be used for assessing the performance in cases where the analytic solution is not known. Secondly, for coupled differential equations with initial conditions, we find that the stability of the solution can be improved by iteratively training on increasing domain sizes. Finally, for the calculation of phase transitions, we employ a two-step training where initially the boundaries are chosen to be the true and false vacua, before the correct boundary conditions are used in the second training. This prevents the network from finding the trivial solution where the field is always in the false vacuum.

\subsection{Ordinary differential equation examples}

To show how well the method can solve ordinary differential equations (ODEs), we apply it to both a first and a second order ODE, which have known analytic solutions. The equations we study are,
\begin{equation}
\frac{d\phi}{dx} + \left (x + \frac{1+3x^2}{1+x+x^3} \right ) \phi - x^3 - 2x - x^2 \frac{1+3x^2}{1+x+x^3}=0~,
\label{eq:firstODE}
\end{equation}
with the boundary condition $\phi(0)=1$ in the domain $x\in [0,2]$ and,
\begin{equation}
\frac{d^2\phi}{dx^2}+ \frac{1}{5} \frac{d\phi}{dx} + \phi + \frac{1}{5} e^{-\frac{x}{5}}\cos x = 0~,
\label{eq:secODE}
\end{equation}
with boundary conditions $\phi(0)=0$ and $\frac{d}{dx} \phi(0) = 1$ in the domain $x \in [0,2]$.

\begin{figure}[t]
\includegraphics[width=1.0\columnwidth]{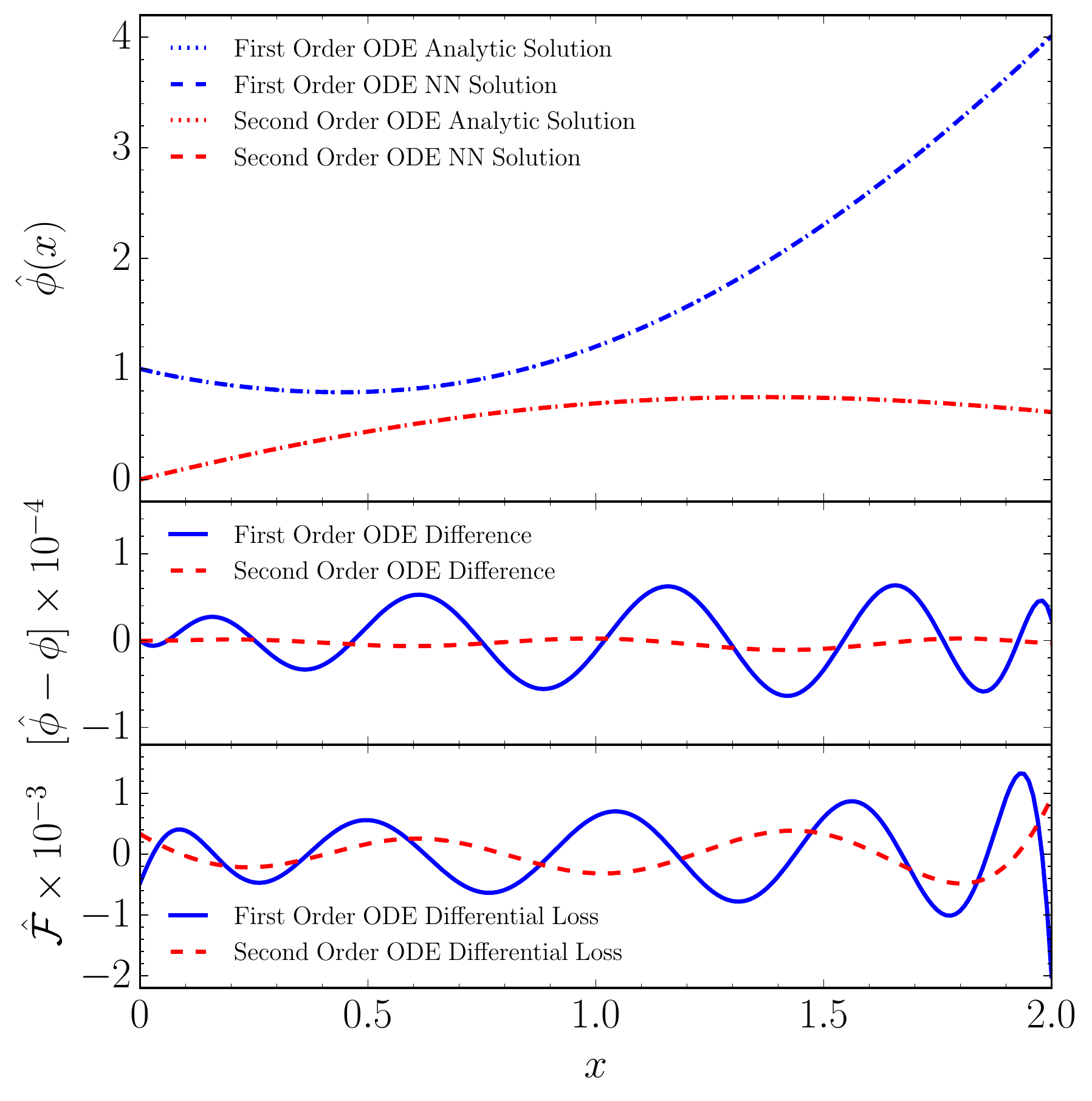}
\caption{The upper panel shows the solutions to the first and second order ODEs of Eqs.~\eqref{eq:firstODE} and \eqref{eq:secODE}, with boundary conditions as outlined in the text. The middle panel shows the numerical difference between the analytic solution and the NN predicted solution for both cases. The lower panel shows the differential contribution $\hat{\mathcal{F}}$ to the loss across the entire domain.}
\label{fig:first_and_second_order_ODEs}
\end{figure}

As a simple neural network structure, we choose a single hidden layer of 10 units with sigmoid activation functions, and we discretise the domain into 100 training examples.  It is then just a case of passing the differential equations and boundary conditions to the loss function, as described in Eq.~\eqref{eq:general_loss_function}, and proceeding with the optimisation. Fig.~\ref{fig:first_and_second_order_ODEs} shows the results of the neural network output, compared to the analytic solutions of Eqs.~\eqref{eq:firstODE} and \eqref{eq:secODE}. The middle panel of Fig.~\ref{fig:first_and_second_order_ODEs} shows the absolute numerical difference between the numerical and analytic solutions. This difference can be reduced further by increasing the number of epochs, the number of anchors in the discretisation of the domain, or the number of units in the hidden layer. Thus, the neural network provides handles to consistently improve the numerical accuracy one aims to achieve. 

The lower panel of Fig.~\ref{fig:first_and_second_order_ODEs} shows the differential contribution to the loss function, i.e. how much each training example contributes to the loss. As we will describe in the next section, if the solution is not analytically known, this provides a measure to assess whether the found solution is the correct one or if a numerical instability led the network to settle in a local minimum for the loss.

\subsection{Coupled differential equation example}

When discussing the calculation of cosmological phase transitions, we will study the solution of coupled non-linear differential equations, for which no closed analytic form is known. Here, we will first show that such solutions can be obtained with our approach, for a case where an analytic solution is known. We consider, 
\begin{align}
&\frac{d\phi_1}{dx} - \cos x - \phi_1^2 - \phi_2 + 1 + x^2 + \sin^2x = 0~,  \nonumber \\
&\frac{d\phi_2}{dx} - 2x + (1+x^2)\sin x - \phi_1\phi_2 = 0~,
\label{eq:coupledDE}
\end{align}
with boundary conditions,
\begin{equation}
\phi_1(0)=0~,~~~\phi_2(0)=1~.
\end{equation}

\begin{figure}[t]
\includegraphics[width=1.0\columnwidth]{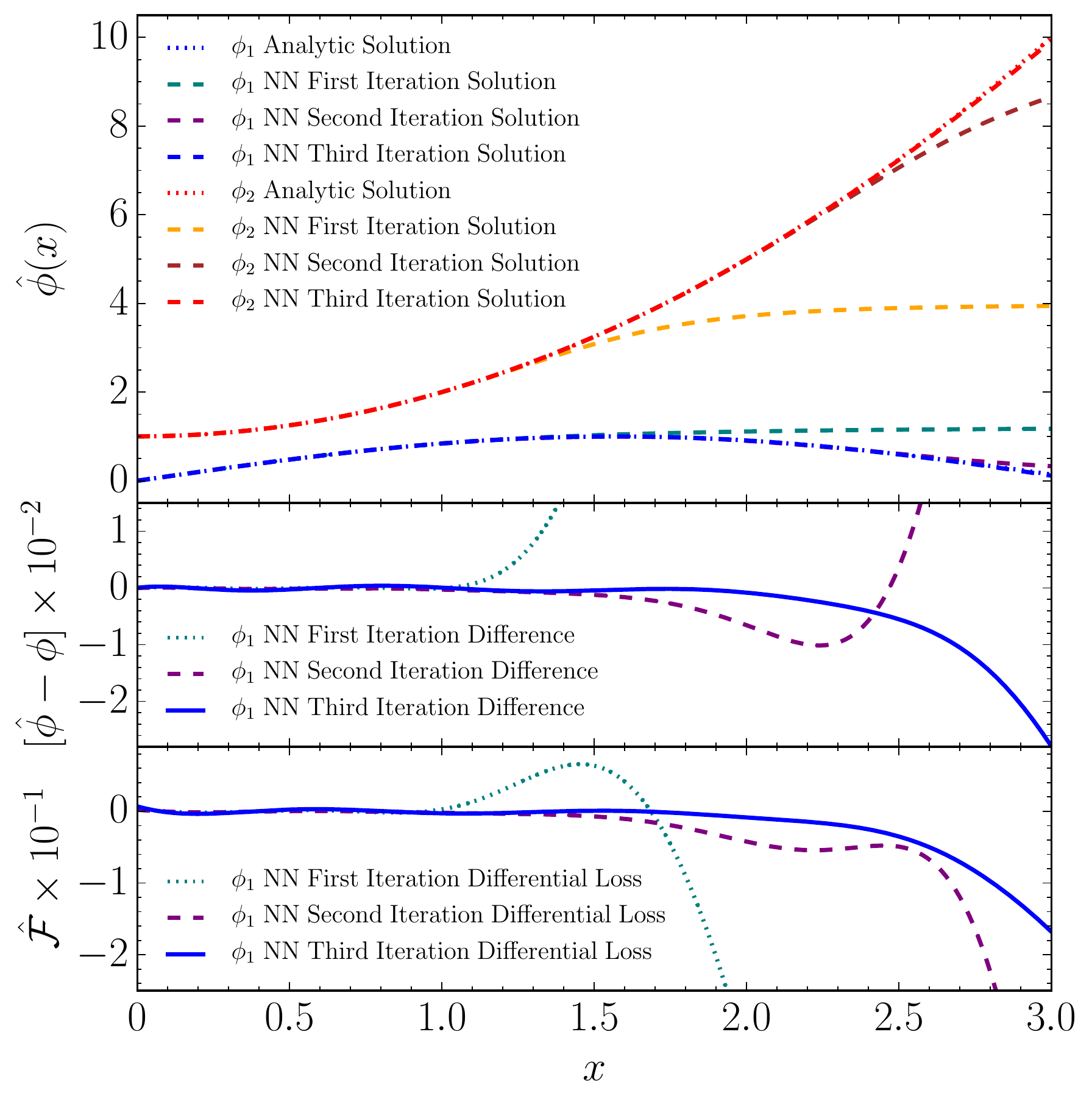}
\caption{The upper panel shows the solutions for the functions $\phi_1$ and $\phi_2$ to the coupled differential equation of Eq.~\eqref{eq:coupledDE}. The middle panel displays the numerical difference between the analytic solution and the NN predicted solution for $\phi_1$. The lower panel shows the differential contribution $\hat{\mathcal{F}}$ to the loss across the entire domain, from the equation for $\phi_1$. The three NN curves in each panel correspond to the first, second and third iteration steps in the training of the network, with iterative increase of the training domain, as described in the text.}
\label{fig:coupled_ODEs}
\end{figure}

If the boundary conditions are set on one end of the domain, e.g. here at $x=0$, it requires an increasingly elaborate network to maintain numerical stability for the solution over a large domain, e.g. where $x \gg 1$. This is due to small numerical instabilities during backpropagation because of the complexity of the loss hypersurface. If such numerical instability leads the network to choose a path that is in close proximity to the true solution, the NN can settle on a local minimum with a small value for the loss function. To solve this problem, we propose to incrementally extend the domain on which a solution should be found, by partitioning the training examples and increasing the number of partitions the NN is trained on in each step. If the weights the NN has learned in the previous step are then retained before training for the next step---i.e. the network only has to learn the function on the part of the domain that was incrementally increased---we find that one can achieve numerical stability for an arbitrarily large domain.

We show this mechanism in Fig.~\ref{fig:coupled_ODEs}, where we have partitioned the full domain containing 100 training examples into three regions each of size 1. The network structure again consists of a single hidden layer of 10 units with sigmoid activation functions, and with two units in the final layer, since there are two coupled equations. The upper panel shows the solutions for $\phi_1$ and $\phi_2$ for each iterative step. While the first iteration only allows a solution to be found on a smaller domain, i.e. here from 0 to 1, subsequent steps, and in particular the third step, allow an accurate solution to be found over the entire domain. Again, the differential $\hat{\mathcal{F}}$ proves to be a good indicator of whether the calculated solution is satisfying the differential equation over the entire domain (see the lower panel of Fig.~\ref{fig:coupled_ODEs}).

\subsection{Partial differential equation example}

\begin{figure}[t]
\includegraphics[width=1.0\columnwidth]{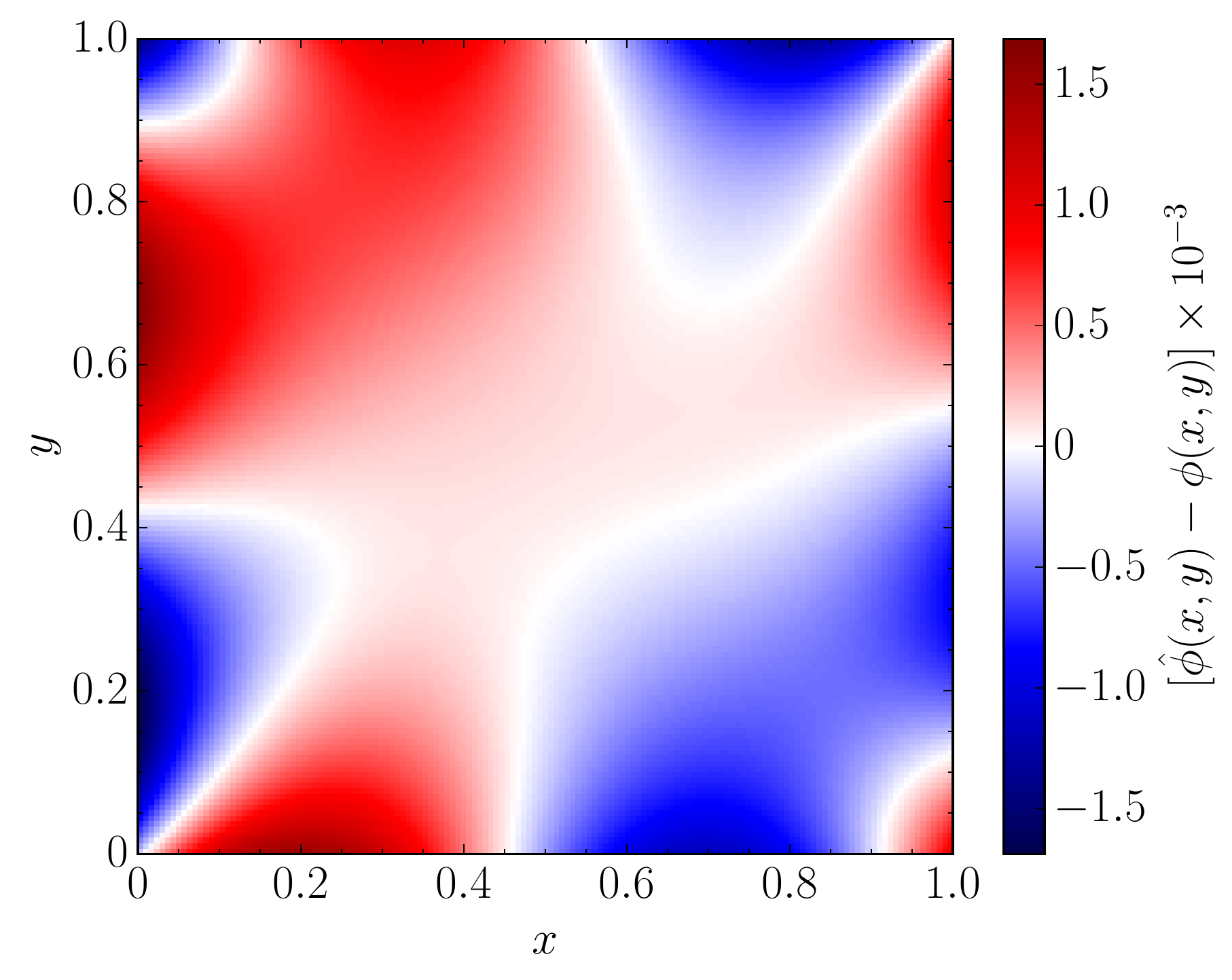}
\caption{Numerical difference between the analytic solution and the NN predicted solution of Eq.~\eqref{eq:PDE}, with boundary conditions as given in Eq.~\eqref{eq:boundPDE}, over the domain $(x,y) \in [0,1] \times [0,1]$.}
\label{fig:second_order_PDE}
\end{figure}

While we do not study partial differential equations (PDEs) in the later physics examples of calculating phase transitions, we showcase here the flexibility of our NN method. With the same network architecture as used for the solution of the ordinary differential equations (except for an extra input unit for each additional variable), we can apply our approach to the solution of partial differential equations. The precise solution of such equations is a widespread problem in physics, e.g. in mechanics, thermodynamics and quantum field theory. As an example, we choose the second order partial differential equation,
\begin{equation}
\nabla^2\phi - e^{-x}(x-2+y^3+6y) = 0~,
\label{eq:PDE}
\end{equation}
with boundary conditions,
\begin{align}
& \phi(0,y)=y^3~,~~~\phi(1,y)=(1+y^3)e^{-1}~,\nonumber \\
& \phi(x,0)=xe^{-x}~,~~~\phi(x,1) = e^{-x}(x+1)~,
\label{eq:boundPDE}
\end{align}
for which an exact analytic solution is known. In Fig.~\ref{fig:second_order_PDE} we show the difference between the numerical solution as predicted by the NN and the analytic solution over the domain $(x,y) \in [0,1] \times [0,1]$. The 100 training examples were chosen from an evenly spaced $10\times10$ grid. As the value of $\phi(x,y)$ is of $\mathcal{O}(1)$ for most of the domain, the relative and absolute accuracies are similar, so we only show the absolute accuracy here. Across the entire domain, we find a numerical solution with very good absolute and relative accuracy for this second order partial differential equation. However, with a deeper NN, e.g. a second layer with 10 tanh units, we find that the accuracy improves by an order of magnitude further. Deeper and wider networks result in even better accuracies.

\section{Calculation of Phase Transitions During the Early Universe}
\label{sec:pt}

Electroweak baryogenesis is a candidate for solving the baryon asymmetry puzzle, the observed abundance of matter over antimatter in the Universe \cite{Shaposhnikov:1986jp, Kuzmin:1985mm}. The need for a dynamical generation of baryon asymmetry is dictated by inflation---the entropy production occurring in the reheating phase of inflation would have washed out any asymmetry already present at the beginning of the Universe's evolution \cite{Linde:2005ht}. A model of baryogenesis was proposed by Sakharov in 1967 \cite{Sakharov:1967dj}, and must now be accommodated by any fundamental theory capable of addressing the baryon asymmetry problem. This is commonly translated into three necessary conditions: (i) baryon number violation, (ii) C- and CP-violation, and (iii) loss of thermal equilibrium. While the first condition can be satisfied in the SM, the second and third conditions require it to be extended \cite{Anderson:1991zb, Dine:1992vs, Huet:1994jb}. Departure from thermal equilibrium can be obtained during a strong first-order phase transition, which is usually accompanied by a sudden change of symmetry \cite{Pathria}. Within the SM, this could have occurred during electroweak symmetry breaking when the Universe had the temperature $T \sim 100 $ GeV \cite{Rubakov:1996vz, Morrissey:2012db}. In order to assess whether this might have been the case, it is crucial to discuss the conditions for scalar-field phase transitions at finite temperature. \\

Quantum fluctuations allow the transition between two vacua of the potential $V(\vec{\phi})$.\footnote{Without loss of generality we consider an $n$-dimensional real scalar field $\vec{\phi}(x)$.} When these are not degenerate, the configuration which corresponds to a local minimum, the false vacuum $\vec{\phi}_{F}$, becomes unstable under barrier penetration, and can decay into the true vacuum $\vec{\phi}_{T}$ of the potential. The tunnelling process converts a homogeneous region of false vacuum into one of approximate true vacuum---a bubble. Far from this region the false vacuum persists undisturbed \cite{Coleman:1977py}. The Euclidean action for this process reads,
\begin{equation}
\mathcal{S}_{4}(\vec{\phi}) = \int d \tau \hspace{2pt} d^{3}  x \hspace{2pt} \Big{[} \frac{1}{2}  \Big{(} \frac{d \vec{\phi} }{d \tau} \Big{)}^{2} +  \frac{1}{2}  ( \nabla \vec{\phi})^{2} + V(\vec{\phi})  \Big{]}~.
\end{equation}

The description of the tunnelling action at finite temperatures follows from the equivalence between the quantum statistics of bosons (fermions) at $T \neq 0$ and Euclidean quantum field theory, periodic (anti-periodic) in the Euclidean time $\tau$ with period $T^{-1}$.  In the calculation of $\mathcal{S}_{4}(\vec{\phi})$, the integration over $\tau$ is replaced  by multiplication by $T^{-1}$ \cite{Linde:1981zj}, leaving the three-dimensional Euclidean action,
\begin{equation}
\mathcal{S}_{3}(\vec{\phi}) = \int d^{3} x \hspace{2pt} \Big{[}  \frac{1}{2} ( \nabla \vec{\phi})^{2} + V(\vec{\phi}, T)  \Big{]}~,
\label{eq:action}
\end{equation}
with $\mathcal{S}_{4}(\vec{\phi}) = T^{-1} \mathcal{S}_{3}(\vec{\phi})$. Suggested by the symmetry of the physical problem, we assume $\vec{\phi}(\vec{x})$ to be invariant under three-dimensional Euclidean rotations (see Ref.~\cite{Coleman:1977th} for a rigorous demonstration in the case of one single scalar field), and  define $\rho = \sqrt{{\vec x}^2}$. The bubble configuration $\vec{\phi}_{b}(\rho)$ is the solution to the Euler-Lagrange equation of motion,
\begin{equation}
\frac{d ^{2} \vec{\phi}}{d \rho^{2}} + \frac{2}{\rho} \hspace{2pt} \frac{d \vec{\phi}}{d \rho} = \nabla V~,
\label{eq:eom}
\end{equation}
where the gradient of the potential is with respect to the field $\vec{\phi}$. The boundary conditions are,
\begin{equation}
\frac{d}{d\rho}\vec{\phi}(0) = 0~,~~~  \lim_{\rho\to\infty} \vec{\phi}(\rho) = \vec{\phi}_{F}.
\label{eq:boundary}
\end{equation}
The solution thus minimises the action. The probability per unit time and unit volume for the metastable vacuum to decay is given by,
\begin{equation}
\frac{\Gamma}{V} = A \hspace{1pt} e^{- B/T}~.
\label{eq:gamma}
\end{equation}
This is maximised by the bounce,
\begin{equation}
B = \mathcal{S}_{3}(\vec{\phi}_{b}) - \mathcal{S}_{3}(\vec{\phi}_{F})~,
\end{equation}
where $\mathcal{S}_{3}(\vec{\phi}_{F})$ is the action evaluated at the stationary configuration $\vec{\phi}_{F}$. A complete expression for the factor $A$ in Eq.~\eqref{eq:gamma} would require complex computations of differential operator determinants, for which we refer the reader to Ref.~\cite{Linde:2005ht}. An estimate can be obtained from dimensional analysis, which gives $A  \sim T^{4}$ \cite{Apreda:2001us}.

Dedicated methods for calculating the nucleation rate, by finding a solution for the bubble profile $\vec{\phi}$ to the non-linear coupled differential equations of Eq.~\eqref{eq:eom}, exist and have been implemented in publicly available codes, e.g. \texttt{CosmoTransitions} \cite{Wainwright:2011kj} and \texttt{BubbleProfiler} \cite{Athron:2019nbd}. For the single-field case, both \texttt{CosmoTransitions} and \texttt{BubbleProfiler} use variants of the overshooting and undershooting method. In the multiple-field case, \texttt{BubbleProfiler} applies the Newton-Kantorovich method \cite{NKtheorem}, as described in \cite{Akula:2016gpl}. \texttt{CosmoTransitions} instead uses a method that splits the equation of motion into a parallel and perpendicular component along a test path through field space. Then the path is varied until a configuration is found that simultaneously solves both directions of the equations of motion. A further code to calculate the tunnelling rates is given in Ref.~\cite{Masoumi:2016wot}. An approach using neural networks to directly learn bounce actions from potentials was described in Ref.~\cite{Jinno:2018dek}. Recently, a novel approximative approach for single \cite{Espinosa:2018hue} and multiple fields \cite{Espinosa:2018szu} was proposed, and a new method based on exact analytic solutions of piecewise linear potentials is outlined in Ref.~\cite{Guada:2018jek}. Older numerical approaches to calculating bubble profiles and tunnelling rates include Refs.~\cite{John:1998ip,Cline:1999wi,Konstandin:2006nd,Park:2010rh}.

\subsection{Phase transition with a single scalar field}

As a first application of our method to the computation of cosmological phase transitions, we consider the case of a  single scalar field. Eq.~\eqref{eq:eom} then has a straightforward classical analogy---it describes the motion of a particle with coordinate $\phi(\rho)$ subject to the inverted potential $-V(\phi)$ and to a peculiar looking damping force which decreases with time. The problem reduces to finding the initial position $\phi_{0}$, in the vicinity of $\phi_{T}$, such that the particle stops at $\phi_{F}$ as $\rho \rightarrow \infty$.

\begin{figure}[t]
\includegraphics[width=1.0\columnwidth]{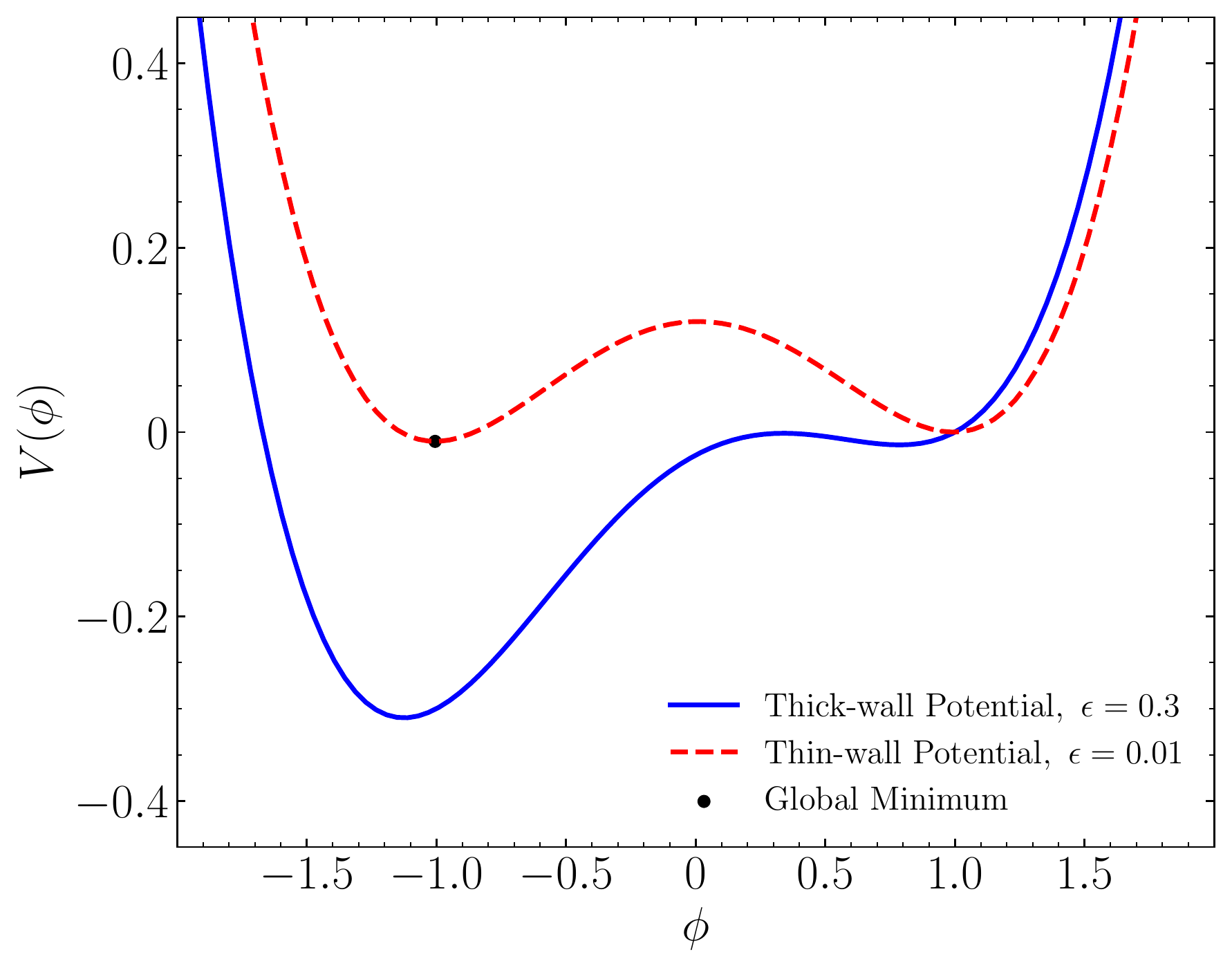}
\caption{Plot of the potential in Eq.~\eqref{eq:1Dpot}, with $\lambda = \alpha = 1$, for the thick-wall (blue solid) and the thin-wall (red dashed) cases. For the latter, the position of the global minimum is also marked by the black dot for clarity.}
\label{fig:1dimPot}
\end{figure}

\begin{figure}[t]
\includegraphics[width=1.0\columnwidth]{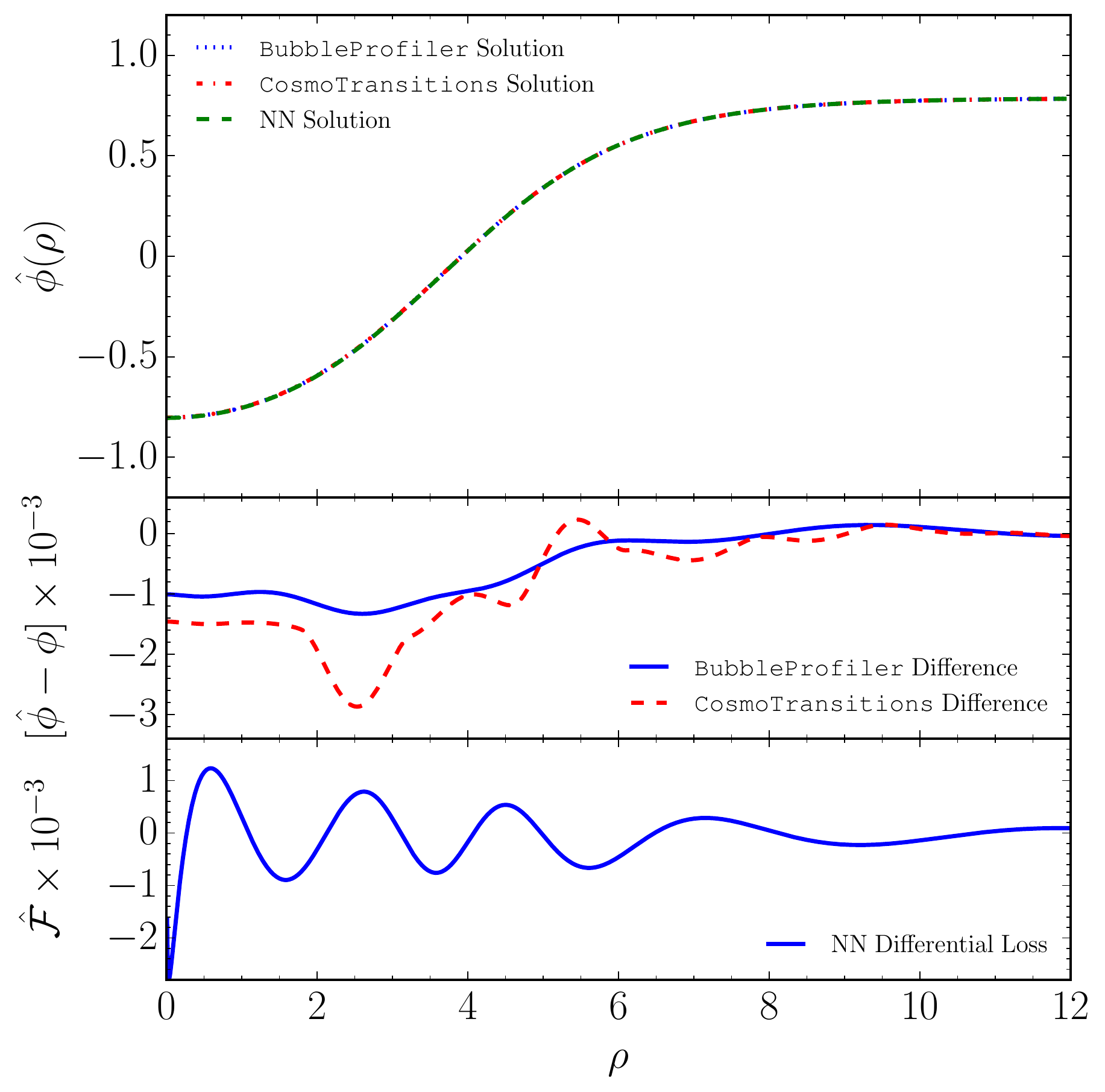}
\caption{The upper panel shows the bubble profile for the thick-wall potential ($\epsilon=0.3$) in Eq.~\eqref{eq:1Dpot} for one scalar field, as obtained by our NN method, \texttt{BubbleProfiler} and \texttt{CosmoTransitions}. The middle panel displays the numerical difference between the NN predicted solution and the solutions from the other two codes. The lower panel shows the differential contribution $\hat{\mathcal{F}}$ to the loss.}
\label{fig:1D_thick_wall}
\end{figure}

\begin{figure}[t]
\includegraphics[width=1.0\columnwidth]{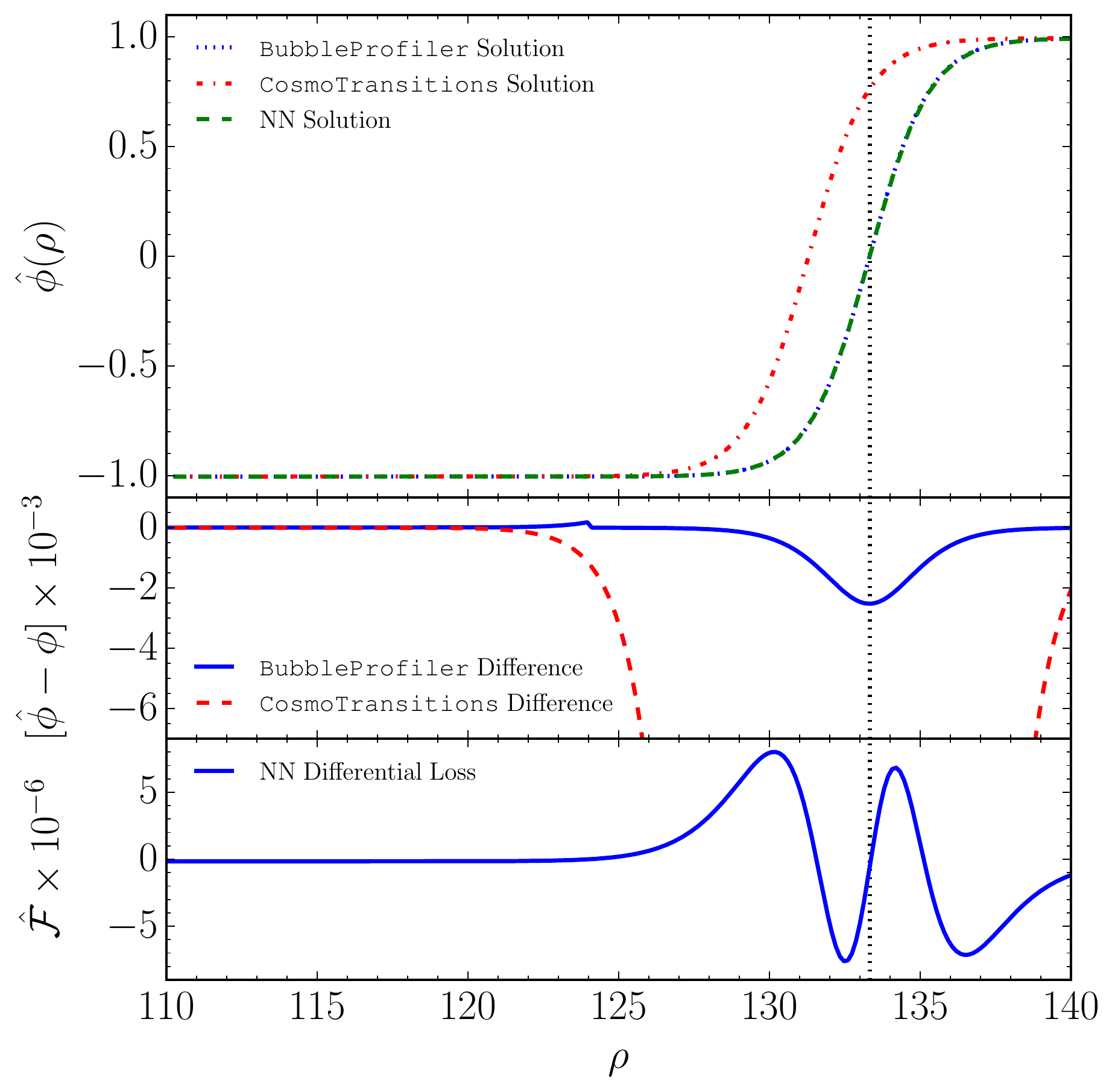}
\caption{The upper panel shows the bubble profile for the thin-wall potential ($\epsilon=0.01$) in Eq.~\eqref{eq:1Dpot} for one scalar field, as obtained by our NN method, \texttt{BubbleProfiler} and \texttt{CosmoTransitions}. The middle panel displays the numerical difference between the NN predicted solution and the solutions from the other two codes. The lower panel shows the differential contribution $\hat{\mathcal{F}}$ to the loss. The dotted vertical line shows the analytic location of the bubble radius, which agrees with the radius found by both the NN and \texttt{BubbleProfiler}.}
\label{fig:1D_thin_wall}
\end{figure}

Existence of a solution was proven in Ref.~\cite{Coleman:1977py}. Starting too close or too far from $\phi_{T}$ would lead to missing the final configuration $\phi_{F}$, due to overshooting and undershooting, respectively. Continuity of $\phi(\rho)$ thus implies that there must exist an intermediate initial position $\phi_{0}$ which solves the boundary conditions in Eq.~\eqref{eq:boundary}. The solution presents two limiting profiles, determined by the ratio of $\Delta \equiv V(\phi_{F}) - V(\phi_{T})$ to the height of the potential barrier $V(\phi_{\mathrm{bar}})$. If this ratio is $\gtrsim 1$, which corresponds to the thick-wall case, the particle will overshoot unless its initial energy is similar to $V(\phi_{F})$. Conversely, if this ratio is small, corresponding to the thin-wall case, in order to avoid undershooting, the particle must wait close to $\phi_{T}$ until the time $\rho \approx R$, when the damping force has become negligible. The value of $R$ can be determined exactly in the thin-wall limit \cite{Coleman:1977py} using,
\begin{equation}
R = \frac{2 \sigma}{\Delta}~,
\label{eq:radius}
\end{equation}
where the surface tension $\sigma$ is given by,
\begin{equation}
\sigma = \lim_{\Delta \rightarrow 0} \int_{\phi_{F}}^{\phi_{T}} d\phi \sqrt{2 [V(\phi) -V(\phi_{F})] }~.
\end{equation}

We test our method on the potential \cite{Athron:2019nbd},
\begin{equation}
V(\phi) = \frac{\lambda}{8}(\phi^2-a^2)^2 + \frac{\epsilon}{2a}(\phi-a)~,
\label{eq:1Dpot}
\end{equation}
and set $\lambda = a = 1$. Two distinct and non-degenerate minima exist for $0 < \epsilon \lesssim 0.3$, with the upper bound representing the thick-wall limit and smaller values of $\epsilon$ representing progressively thinner cases. A plot of the potential is shown in Fig.~\ref{fig:1dimPot}, for the values $\epsilon=0.01$ and $\epsilon=0.3$ which we consider as our thin-wall and thick-wall cases, respectively.

For the boundary conditions in Eq.~\eqref{eq:boundary}, it is clearly not possible to implement an infinite domain for the training of a neural network, and the divergence in the second term of Eq.~\eqref{eq:eom} prevents the equation from being evaluated at $\rho=0$. Therefore, a training domain $\rho \in [\rho_{\mathrm{min}}, \rho_{\mathrm{max}}]$ must be chosen. Since the solution approaches the boundaries exponentially, it can be safely assumed that the numerical uncertainties induced by these choices can be neglected, provided that $\rho_{\mathrm{min}}$ is sufficiently small and $\rho_{\mathrm{max}}$ is sufficiently large. To help in choosing this domain, the identification of $\epsilon$ in Eq.~\eqref{eq:1Dpot} with $\Delta$ in Eq.~\eqref{eq:radius} can be made, and the bubble radius $R$ calculated. We then use $\rho_{\mathrm{max}}=5R$ for the thick-wall case, and $\rho_{\mathrm{max}}=2R$ for the thin-wall case (since Eq.~\eqref{eq:radius} underestimates the true radius for thick-wall cases). Furthermore, we use $\rho_{\mathrm{min}}=0.01$ for both cases. Although these choices may seem arbitrary, we find that the solution converges provided that the transition point is contained well inside the domain, and the result remains stable even if larger domains are used. The boundary conditions then read,
\begin{equation}
\frac{d}{d\rho}\phi(\rho_{\mathrm{min}}) = 0~, ~~~ \phi(\rho_{\mathrm{max}}) = \phi_{F}~.
\label{eq:boundary2}
\end{equation}

Our NN method can then be applied to find the bubble profile by solving the Euler-Lagrange equation in Eq.~\eqref{eq:eom}.
In this context, the NN method corresponds to an approach where the neural network attempts to apply the minimum action principle to the Euclidean action of Eq.~\eqref{eq:action}. The test-field configuration, defined by the output layer of the neural network, is then adjusted using backpropagation until the classical trajectory is found. We discretise the domain into 500 training points and choose a network with a single hidden layer. For the thick-wall case, we use 10 hidden units with a sigmoid activation function, as was used in earlier examples; however, for the thin-wall case, we find that a single tanh unit is sufficient to achieve very good performance since the solution itself closely resembles a tanh function. To prevent the network from finding the trivial solution where the field remains in the false vacuum forever, we first train the network with the boundary condition at $\rho_{\mathrm{min}}$ modified to $\phi(\rho_{\mathrm{min}})=\phi_{T}$ so that the network finds a solution in the vicinity of the correct solution, since the starting point is close to the true vacuum, before training again with the correct boundary conditions. We use this two-step training for all phase transition calculations.

Our results for the thick-wall and thin-wall cases are shown in Figs.~\ref{fig:1D_thick_wall} and \ref{fig:1D_thin_wall}, respectively, together with the \texttt{CosmoTransitions} and \texttt{BubbleProfiler} solutions. While all three methods agree very well for the thick-wall case, there is a disagreement in \texttt{CosmoTransitions} compared to \texttt{BubbleProfiler} and the NN approach in the thin-wall case. The dotted vertical line indicates where the bubble radius should be according to Eq.~\eqref{eq:radius}. Both, \texttt{BubbleProfiler} and NN find a solution that matches the analytic calculation for the bubble radius. \texttt{CosmoTransitions} instead finds a solution with a smaller bubble radius, and therefore a smaller action and a larger tunnelling rate. 

For thin-wall cases, numerical stability is difficult to achieve. It is possible for an approximate solution to be found, which transitions at a much earlier $\rho$ than it should, since a translated solution also approximately solves the differential equation \cite{Masoumi:2016wot}. For our method, $\hat{\mathcal{F}}$ can be monitored during the course of the training. During the early stages of the training where the solution does not yet have the correct transition point, $\hat{\mathcal{F}}$ will be sharply distributed in the region of the incorrect transition. As the training proceeds and the solution converges, the function will flatten out until an accurate solution is found across the entire domain.

We have shown that the NN achieved very good stability for the thin-wall case using a single tanh function. We also explored the idea of using an adaptive distribution of training examples, such that more examples are distributed close to the region where the transition of the NN solution happens, and this distribution is then modified over the course of the training. A larger contribution to the loss in this region will be amplified by having more training examples, which can speed up learning. We found that the results can be improved by using this procedure, and this is an idea which could be investigated further in future work.

\subsection{Phase transition with two scalar fields}

\begin{figure}[t]
\includegraphics[width=1.0\columnwidth]{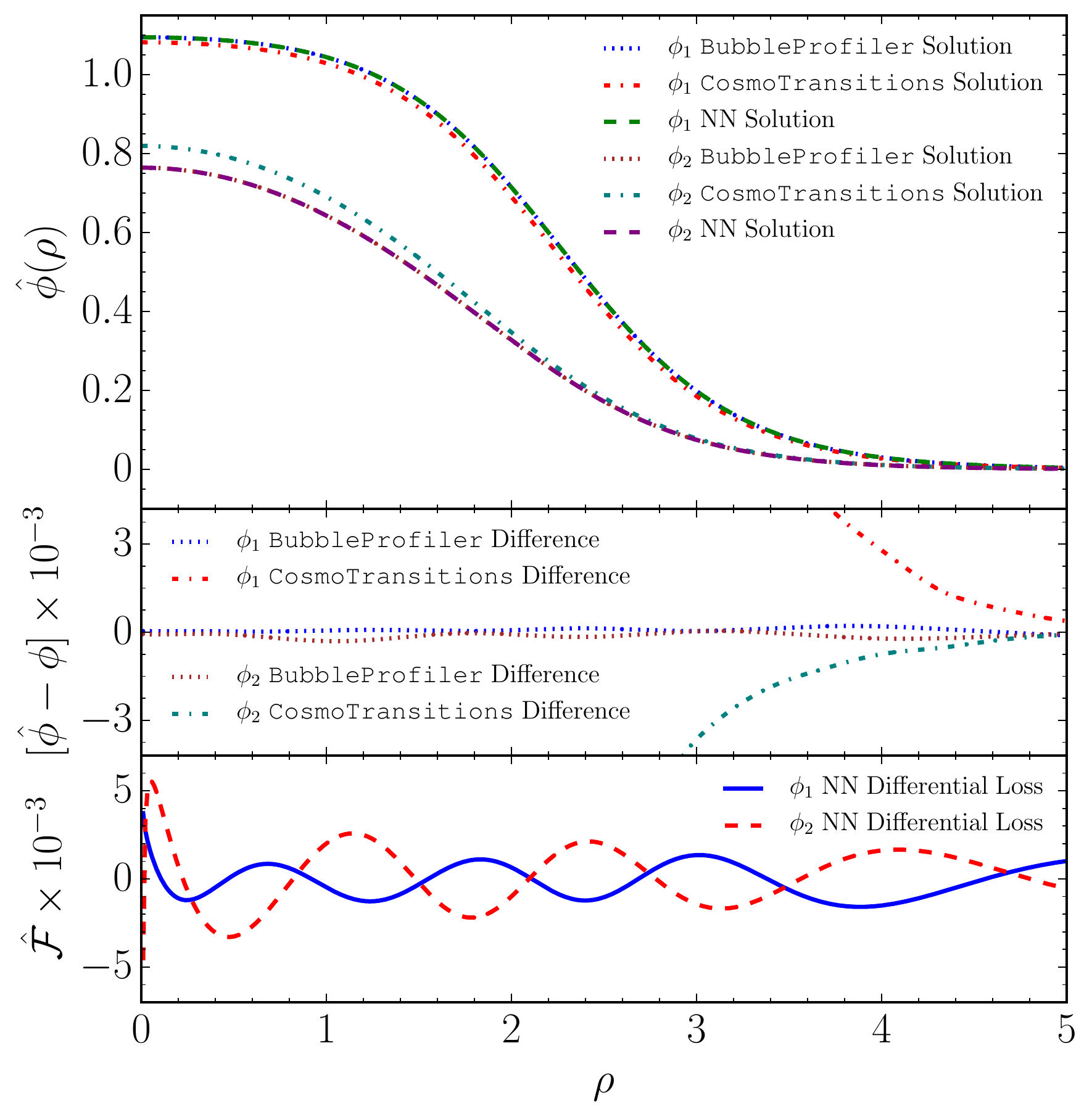}
\caption{The upper panel shows the bubble profiles for the thick-wall potential in Eq.~\eqref{eq:2Dpot} for two scalar fields, as obtained by our NN method, \texttt{BubbleProfiler} and \texttt{CosmoTransitions}. The middle panel displays the numerical difference between the NN predicted solutions and the solutions from the other two codes. The lower panel shows the differential contribution $\hat{\mathcal{F}}$ to the loss from $\phi_1$ and $\phi_2$.}
\label{fig:2D_thick_wall}
\end{figure}

\begin{figure}[t]
\includegraphics[width=1.0\columnwidth]{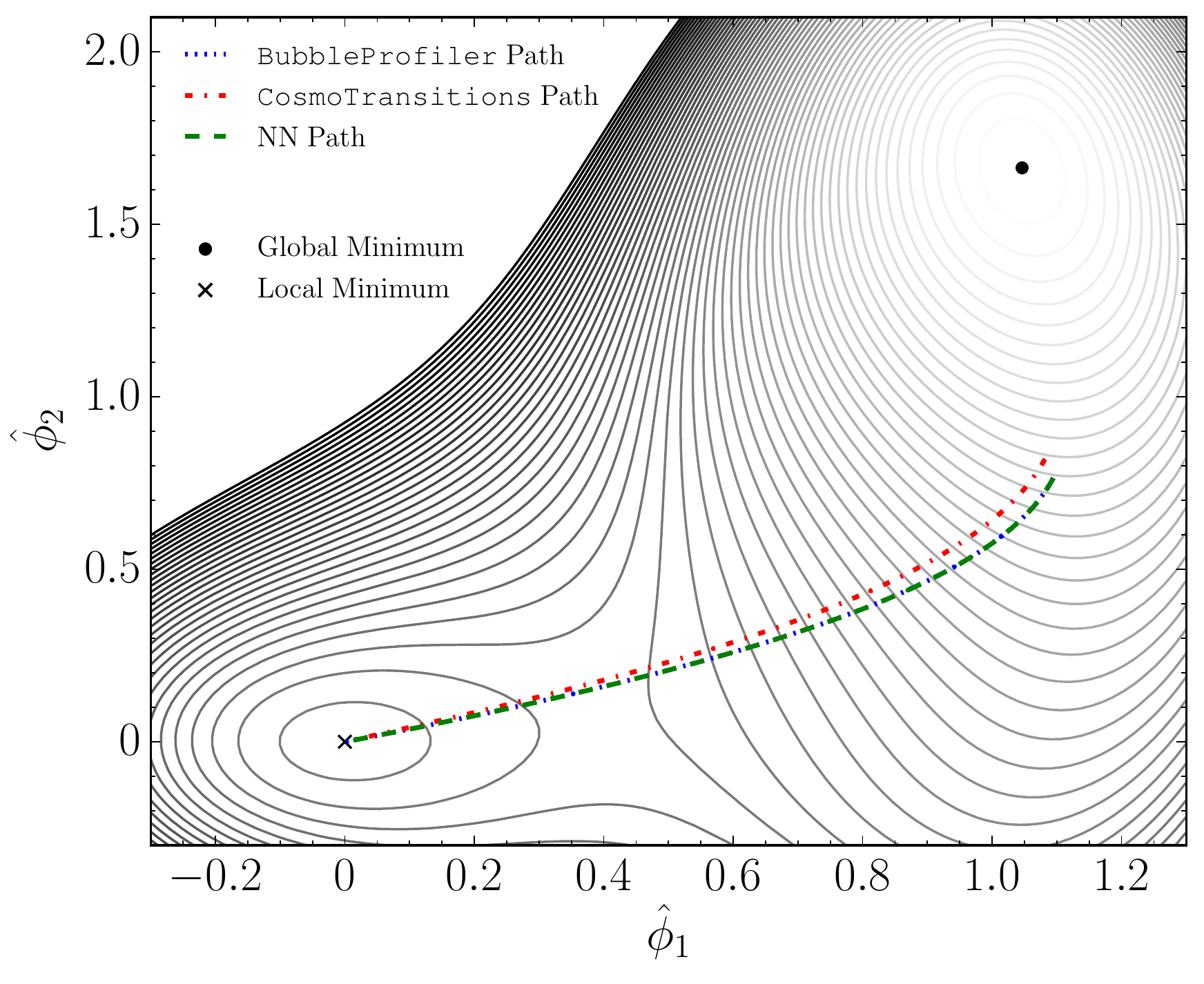}
\caption{Calculated solutions for the tunnelling path for NN, \texttt{BubbleProfiler} and \texttt{CosmoTransitions}. The paths range from the local minimum to the exit point of the tunnelling barrier. Also shown are the contours of the potential, where the global minimum is denoted by the black dot and the local minimum by the black cross.}
\label{fig:2D_thick_wall_path}
\end{figure}

To investigate how well the NN approach can solve the differential equation of Eq.~(\ref{eq:eom}) for multiple fields, we consider a potential for two scalar fields \cite{Wainwright:2011kj},
\begin{equation}
\label{eq:2Dpot}
V(\phi_1,\phi_2) = (\phi_1^2+\phi_2^2) \left [ \frac{9}{5}(\phi_1-1)^2+\frac{1}{5}(\phi_2-1)^2-\delta \right ],
\end{equation}
which has a local minimum at $\phi_1=\phi_2=0$ and a global minimum near $\phi_1 \approx \phi_2 \approx1$. We focus again on the thick- and thin-wall cases, setting $\delta = 0.4$ for the former and $\delta = 0.02$ for the latter. For the thick-wall potential, we solve the coupled equations in Eq.~(\ref{eq:eom}) with the boundary conditions,
\begin{align}
&\frac{d}{d\rho}\phi_1(\rho_{\mathrm{min}})=0~, ~~~ \phi_1(\rho_{\mathrm{max}})=0~, \nonumber \\
&\frac{d}{d\rho}\phi_2(\rho_{\mathrm{min}})=0~, ~~~ \phi_2(\rho_{\mathrm{max}})=0 ~,
\end{align}
in the training domain $\rho \in [0.01, 6]$ with 500 training points. Again, the NN is built with $10$ units in a single hidden layer with a sigmoid activation function. Since there are two fields, the NN has two units in the final layer. The two components $\phi_1$ and $\phi_2$ of the bubble solution, and the associated path through field space, are shown in Figs.~\ref{fig:2D_thick_wall} and \ref{fig:2D_thick_wall_path}, respectively. Once more, \texttt{BubbleProfiler} and the NN predictions agree very well, both for the one-dimensional profiles for $\phi_1$ and $\phi_2$, and for the path in the $(\phi_1, \phi_2)$ plane. \texttt{CosmoTransitions} shows a slightly different shape for the solutions of $\phi_1(\rho)$ and even more so for $\phi_2(\rho)$, resulting in a slightly modified escape path in Fig.~\ref{fig:2D_thick_wall_path}. The behaviour and small numerical value of the differential contribution $\hat{\mathcal{F}}$ to the loss suggests that the NN has converged to a correct solution for the profiles. Since it also agrees very closely with the result from \texttt{BubbleProfiler}, we conclude that in this case the \texttt{BubbleProfiler} result is correct. We note that our NN solution has found initial positions for the fields which agree with those from  
\texttt{BubbleProfiler}. In thick-wall cases, these can differ significantly from the true vacuum $\phi_T$---these initial positions have been independently found by the network during optimization and have not been used as an input during training.

For the thin-wall potential we find that the performance can be significantly improved if a deeper network is used. \texttt{BubbleProfiler} instead does not find a solution at all, while the NN agrees very well with the path found by \texttt{CosmoTransitions}. Since there is not a solution from all three codes, we do not show the plot here.

Thus, in this section we have shown examples where \texttt{CosmoTransitions} or \texttt{BubbleProfiler} fail to provide a correct result, while the NN approach can cope well with both the thick-wall and the thin-wall solutions.

\subsection{Singlet-scalar extended Standard Model with finite temperature contributions}

\begin{figure}[t]
\includegraphics[width=1.0\columnwidth]{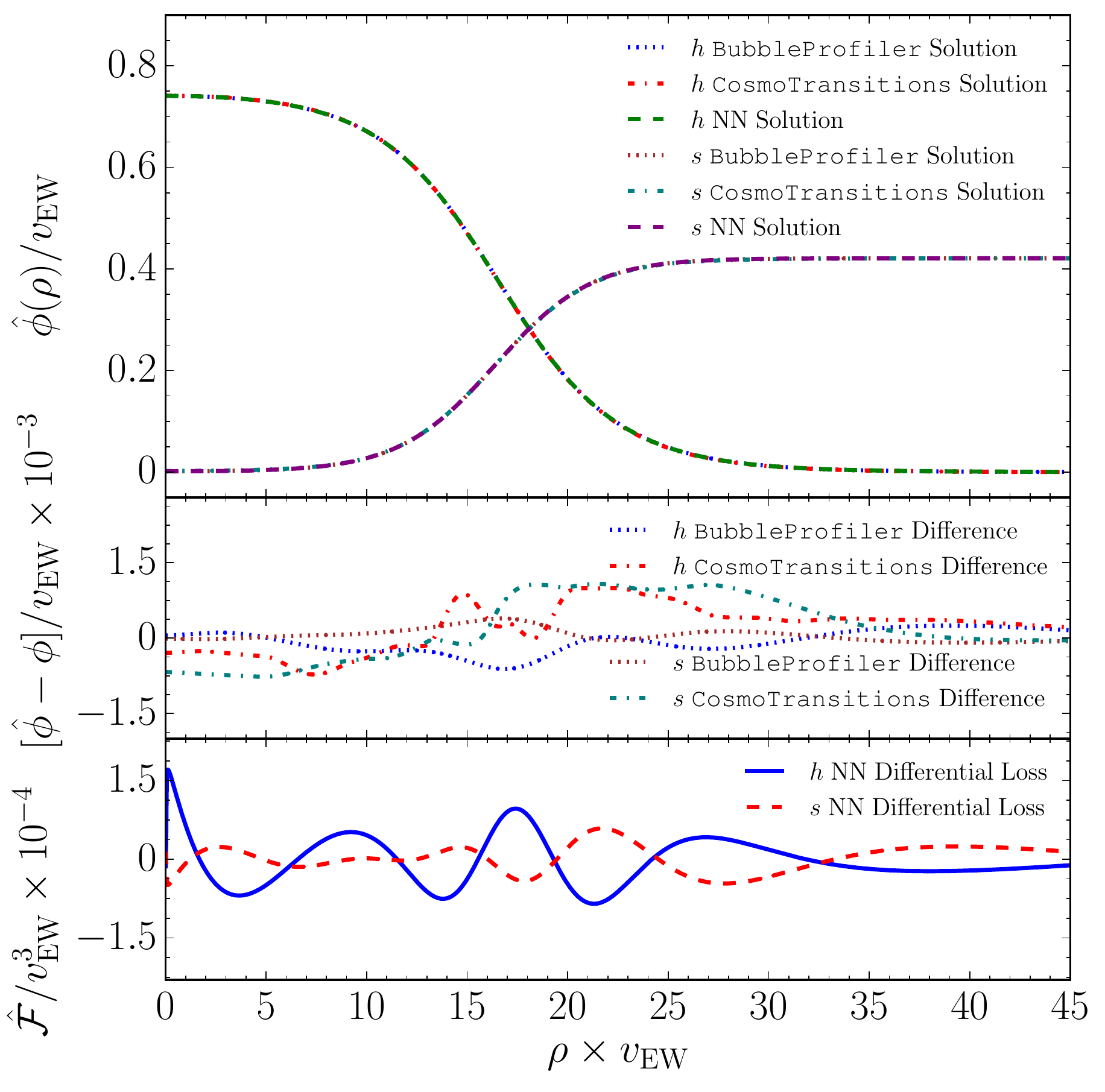}
\caption{The upper panel shows the bubble profiles for the singlet-scalar extended Standard Model potential in Eq.~\eqref{eq:SSM_pot}, as obtained by our NN method, \texttt{BubbleProfiler} and \texttt{CosmoTransitions}. The middle panel displays the numerical difference between the NN predicted solutions and the solutions from the other two codes. The lower panel shows the differential contribution $\hat{\mathcal{F}}$ to the loss from $h$ and $s$.}
\label{fig:SSM}
\end{figure}

As a final example, we study a scenario of phenomenological interest, namely the extension of the SM Higgs sector by a single scalar field.\footnote{For projected and existing limits on this model see Refs.~\cite{Profumo:2007wc,No:2018fev} and references therein.} Despite its simplicity, the singlet-scalar extended Standard Model (SSM) could potentially provide solutions to puzzles such as the existence of dark matter \cite{Kusenko:2006rh, Burgess:2000yq, McDonald:2001vt} and electroweak baryogenesis \cite{ Cline:2012hg, Chala:2016ykx, Huber:2006wf, Huber:2000mg}, where a crucial requirement is a strong electroweak phase transition, as discussed previously. The tree-level potential reads,
\begin{align}
V^{(0)}(h, s) = &-\frac{1}{2} \mu_{h}^{2} h^{2} + \frac{1}{4} \lambda_{h}h^{4} + \frac{1}{2}\mu_{s}^{2}s^{2}  \nonumber\\
&+ \frac{1}{4} \lambda_{s} s^{4} + \frac{1}{4} \lambda_{m} s^{2} h^{2}~,
\label{eq:V0}
\end{align}
where $h$ denotes the Higgs field and $s$ the additional $\mathbb{Z}_{2}$-symmetric scalar field.\footnote{This condition could also be relaxed, since in models with no $\mathbb{Z}_{2}$-symmetry the most general renormalisable potential would have three more parameters \cite{Profumo:2007wc,Espinosa:2011ax}.} It is possible to consider a scenario in which the potential barrier separating the electroweak symmetric and the broken phase is generated already at tree level \cite{Espinosa:2011ax}. In this scenario, to study the evolution of parameters with $T$, it is enough to include only the high-temperature expansion of the one-loop thermal potential, which results in thermal corrections to the mass parameters \cite{Cline:1999wi},
\begin{equation}
V^{(1)}(h, s, T) = \Big{(} \frac{1}{2} c_{h} h^{2} + \frac{1}{2} c_{s} s^{2} \Big{)} T^{2}~,
\label{eq:V1}
\end{equation}
where,
\begin{equation}
c_{h} = \frac{1}{48} \Big{[} 9 g^{2} + 3 g^{\prime \hspace{1pt}2} + 2 (6 h_{t}^{2}+ 12 \lambda_{h} + \lambda_{m}) \Big{]}~,
\end{equation}
\begin{equation}
c_{s} = \frac{1}{12} (2 \lambda_{m} + 3 \lambda_{s})~,
\end{equation}
with $g$ and $g^{\prime}$ being the $SU(2)_{L}$ and $U(1)_{Y}$ gauge couplings, respectively, and $h_{t}$ is the top Yukawa coupling. We then consider Eq.~(\ref{eq:eom}) with the potential,
\begin{equation}
\label{eq:SSM_pot}
V(h,s,T) = V^{(0)}(h,s) + V^{(1)}(h,s, T)~. 
\end{equation}

At high temperatures the thermal contribution in Eq.~\eqref{eq:V1} dominates, and the global minimum is the $\mathbb{Z}_{2}$ and electroweak symmetric configuration $(h = 0, s = 0)$. The behaviour as $T$ decreases is determined by the choice of parameters. These are constrained to the parameter region in which the potential develops a strong tree-level barrier at the critical temperature $T_{C}$ \cite{Espinosa:2011ax}. In particular, at $T > T_{C}$ after $\mathbb{Z}_{2}$-symmetry breaking, $s$ acquires a non-zero vacuum expectation value, $\expval{s} =  w$, along the $\expval{h} = 0$ direction. This configuration constitutes a global minimum for the potential. At $T = T_{C}$ a second degenerate minimum appears at the electroweak symmetry breaking phase $\expval{h} = v$ and at the restored $\mathbb{Z}_{2}$-symmetric vacuum, $\expval{s} = 0$. Finally, at $T< T_{C}$ the electroweak minimum $(v, 0)$ represents the only energetically favourable configuration. The nucleation temperature at which the phase transition from $\vec{\phi}_{F} = (0, w)$ to $\vec{\phi}_{T}= (v, 0)$ occurs is found from the requirement $\mathcal{S}_{3}(T_{N})/T_{N} \simeq 140$ \cite{Anderson:1991zb, Moreno:1998bq}.

As an example parameter configuration, we consider $T_C = 110$ GeV, $\lambda_m = 1.5$ and $\lambda_s = 0.65$, as used in Ref.~\cite{Athron:2019nbd}, and a temperature of $T=85$ GeV, which is the nucleation temperature that \texttt{BubbleProfiler} finds. We thus solve Eq.~\eqref{eq:eom} with the boundary conditions,
\begin{align}
&\frac{d}{d\rho}h(\rho_{\mathrm{min}})=0~, ~~~ h(\rho_{\mathrm{max}})=0~, \nonumber \\
&\frac{d}{d\rho}s(\rho_{\mathrm{min}})=0~, ~~~ s(\rho_{\mathrm{max}})=w ~.
\end{align}
We use a neural network with 10 units in a single hidden layer with a sigmoid activation function, on a training domain of $\rho \in [0.01, 50]$ with 500 training points. To avoid large numerical values in the loss function, we scale all mass parameters in the potential by the electroweak symmetry breaking vacuum expectation value at zero temperature, $v_{\mathrm{EW}}$. Our result, along with the comparison to \texttt{CosmoTransitions} and \texttt{BubbleProfiler}, is shown in Fig.~\ref{fig:SSM}. We find very good agreement between all three methods to calculate the bubble profiles $h(\rho)$ and $s(\rho)$, and the small values of $\hat{\mathcal{F}}$ across the domain show that good convergence has been achieved.

\section{Conclusions}\label{sec:conclusions}

By building on the capabilities of an artificial neural network in solving optimisation problems, we have proposed a novel way to find solutions to differential equations.

Our method extends existing approaches on several accounts: (i) We avoid trial functions by including the boundary conditions directly into the loss function;
(ii) the differential shape of $\hat{\mathcal{F}}$ is an excellent indicator of whether a good solution to $\mathcal{F}$ has been found over the entire domain; (iii) in regions of numerical stability we propose increasing the domain iteratively to find stable solutions over arbitrarily large domains; (iv) for solutions that vary quickly over a small part of the domain, we find that it can be numerically beneficial to self-adaptively distribute more anchors in such regions.

We applied this approach to finding fully differentiable solutions to ordinary, coupled and partial differential equations, for which analytic solutions are known. Various network architectures have been studied, and even relatively small networks showed a very good performance. 

To show how this method can be applied to a task of direct phenomenological interest, we used it to calculate the tunnelling profiles of electroweak phase transitions and compared them to those obtained by \texttt{CosmoTransitions} and \texttt{BubbleProfiler}. We have presented explicit examples where the neural network method finds correct solutions, while either \texttt{CosmoTransitions} or \texttt{BubbleProfiler} fails. We find an optimised neural network to be very flexible and reliable, and is able to converge to solutions for all the examples tested with an accuracy that is competitive with the dedicated programs for calculating the bubble profiles. However, further work, e.g. in developing an approach to choosing the domain sizes for phase transitions in a more robust way, would be required to develop a fully automated tool using this approach.

As this method could be straightforwardly extended beyond the calculation of differential equations, we envision it to be applicable to a wide range of problems in perturbative and non-perturbative quantum field theory.

\vspace{1cm}
\noindent 
\acknowledgments{
\small{
\textit{
Acknowledgements: 
We would like to thank Mikael Chala for help with the \texttt{CosmoTransitions} code.
MS acknowledges the generous hospitality of Barbara Jaeger and her group at the University of Tuebingen and support of the Humboldt Society during the completion of this work.
}}}


\bibliographystyle{apsrev4-1} 
\bibliography{references} 

\end{document}